\documentstyle[12pt]{article}    
\begin{document}           
\baselineskip=0.33333in
\begin{quote} \raggedleft TAUP 2686-2001
\end{quote}
\vglue 0.5in
\begin{center}{\bf Remarks on the Physical Meaning of \\ 
Diffraction-Free Solutions of Maxwell Equations}
\end{center} 
\begin{center}E. Comay 
\end{center}
 
\begin{center}
School of Physics and Astronomy \\
Raymond and Beverly Sackler Faculty of Exact Sciences \\
Tel Aviv University \\
Tel Aviv 69978 \\
Israel
\end{center}
\vglue 0.5in
\vglue 0.5in
\noindent
PACS No: 03.50.De, 41.20.Jb 
\vglue 0.2in
\noindent
Abstract:

   It is proved that a source of electromagnetic radiation cannot emit
a diffraction-free beam at the wave zone. A Bessel
$J_0$ $\varphi $-invariant beam does not hold even at the intermediate
zone. These results negate claims published recently in the literature.
\newpage

An idea of creating a diffraction-free beam has been published[1]. 
The beam's amplitude
is cylindrically symmetric
($\varphi $-invariant) where the $r$-dependence is 
proportional to the Bessel function of the first kind $J_0(ar)$
and $a$ is a constant having the dimension of $L^{-1}$.
Reference [1] has arisen a great interest in utilization of $J_0$
beams and has been cited more than 360 times[2]. An application
of [1] shows the central peak of the 
assumed $J_0$ beam[3] and another one
refers to its peculiar $z$-component wavelength[4]. 
Another publication related to [1]
claims that a superluminal propagation of light in air
has been detected[5]. Objections to [5] have been published[6].
The purpose of this work is to show that one {\em cannot}
construct a diffraction-free electromagnetic beam 
at the wave zone and that the Bessel function $J_0(ar)$ is
unsuitable for describing diffraction free $\varphi $-invariant
wave at the intermediate zone too. This
outcome proves that results of papers discussing this topic, in
general, and those ascribing superluminal velocity to beams that
take the form of Bessel function $J_0$, in particular, should
be reevaluated. Units where the speed of light 
$c=1$ are used. The metric $g_{\alpha \beta}$
is diagonal and its entries are
(1,-1,-1,-1). ${\bf u}_r $, ${\bf u}_\varphi $ and ${\bf u}_z$
denote unit vectors in cylindrical coordinates and ${\bf u}_x$,
${\bf u}_y$ and ${\bf u}_z$ are unit vectors in Cartesian coordinates.

   A general analysis of diffraction-free solutions of Maxwell
equations has been published[7]. Here the 
fields solving the problem are derived
from a vector potential $\bf {A}$ that satisfies
the wave equation together with the Lorentz-gauge requirement[8].
It turns out that this work is relevant to [1] and some of its
results are analyzed here in detail. (Another work[9] is
closely related to [1] and [7].)
Let us start with the solution 
obtained in Example 1 (on p. 1557 of [7]). Using cylindrical 
coordinates and removing constant 
factors, the time dependent monochromatic
electric field of this solution is obtained from the vector
potential ${\bf E} = -\partial {\bf A}/\partial t$
\begin{equation}
{\bf E} = \omega J_1(ar)e^{i(bz-\omega t)} {\bf u}_\varphi 
\label{eq:E1}
\end{equation}
where $J_1$ is the Bessel function of the first kind of order 1.
The magnetic field is ${\bf B}= curl {\bf A}$
\begin{equation}
{\bf B} = -b J_1(ar)e^{i(bz-\omega t)} {\bf u}_r -
              ia J_0(ar)e^{i(bz-\omega t)} {\bf u}_z .
\label{eq:B1}
\end{equation}

   Ignoring constant factors, one finds that the magnetic field
$(\!\!~\ref{eq:B1})$ is dual to the electric field of Example 2 of
[7]. (The factor 2 in $U_r$ of example 2 is a misprint.) This outcome
indicates that Examples 1 and 2 of [7] represent dual 
electromagnetic solutions where 
${\bf E \rightarrow B,\;\; B \rightarrow -E}$ (see [8], p. 252).

   Having the solution, let us examine the problem of a cylindrically
shaped wave guide whose walls are made of a perfect conductor 
(see [8], p. 335). The length of the cylinder is much greater
than both its diameter $2R$ and the wavelength $\lambda = 1/\omega $
(see fig. 1). The boundary
conditions along the wave guide's walls are (see [8], p. 335)
\begin{equation}
E_\parallel = 0,\;\; B_\perp = 0. 
\label{eq:BOUNDARY}
\end{equation}
Thus, the
solution $(\!\!~\ref{eq:E1})$ and $(\!\!~\ref{eq:B1})$ satisfies the
boundary conditions provided 
\begin{equation}
J_1(aR) = 0.
\label{eq:BOUNDARY2}
\end{equation}

   Dynamical properties of the solution $(\!\!~\ref{eq:E1})$ and
$(\!\!~\ref{eq:B1})$ are obtained from the
energy-momentum tensor of the electromagnetic fields (see 
[10], p. 81 or [8], p. 605))
\begin{equation}
T_{F}^{\mu \nu } =
\frac {1}{4\pi }(F^{\mu \alpha }F^{\beta \nu }g_{\alpha \beta }
+\frac {1}{4}F^{\alpha \beta }F_{\alpha \beta }g^{\mu \nu })
\label{eq:TF}
\end{equation}
where $F^{\mu \nu}$ denotes the tensor of the electromagnetic
fields. Expression $(\!\!~\ref{eq:TF})$
is quadratic in the fields. Hence, one should use 
the real part of $(\!\!~\ref{eq:E1})$ and $(\!\!~\ref{eq:B1})$ in
an evaluation of quantities belonging to it. 
Let us first examine the momentum
density of the fields. This is the Poynting vector
\begin{equation}
{\bf S} = \frac {1}{4\pi }{\bf E} \times {\bf B}.
\label{eq:POYNTING}
\end{equation}
   
   The $z$-component of the momentum density and energy flux are
obtained from the substitution of the appropriate real part of
$(\!\!~\ref{eq:E1})$ and $(\!\!~\ref{eq:B1})$
\begin{equation}
S_z = \frac {b\omega }{4\pi }J_1^2(ar)cos^2(bz-\omega t). 
\label{eq:PZ}
\end{equation}
Expression $(\!\!~\ref{eq:PZ})$ is non-negative at all points, a
property which is consistent with the beam's expected flux of energy 
that travels away from a localized source.

   The radial component of the momentum density is obtained analogously
\begin{equation}
S_r = -\frac {a\omega }{8\pi }J_1(ar) J_0(ar) sin[2(kz-\omega t)].
\label{eq:PR}
\end{equation}
Here one sees that, unlike the case of $(\!\!~\ref{eq:PZ})$, the sign
of $(\!\!~\ref{eq:PR})$ alternates periodically in the time and 
$z$-coordinates. Moreover, for any fixed value of $t$ and $z$, it changes
sign along the $r$-axis, because zeroes of the Bessel functions $J_0$
and $J_1$ do not coincide[11]. 
It follows that although the radial motion does
not vanish locally, its mean value is null. This property indicates
that the radial motion takes the type of a standing wave.

   Now let us examine the interaction of the fields with the walls
of the wave guide. Point $P$ at $x=R,\;y=z=0$ is used as a
representation of the general case and cartesian coordinates
are used. The $x$-component of the momentum current at $P$ is
(see [10], p. 82 or [8], p. 605))
\begin{equation}
T_{xx} = \frac {1}{8\pi }
(E_y^2 + E_z^2 - E_x^2 + B_y^2 + B_z^2 - B_x^2).
\label{eq:TXX1}
\end{equation}
Examining the fields $(\!\!~\ref{eq:E1})$ and $(\!\!~\ref{eq:B1})$ 
and the boundary value $(\!\!~\ref{eq:BOUNDARY2})$, one finds that
only the $z$-component of the magnetic field makes a nonvanishing
contribution. Thus, the momentum current at $P$ is 
\begin{equation}
T_{xx} = \frac {a^2}{8\pi }J_0^2(aR)sin^2(bz-\omega t).
\label{eq:TXX2}
\end{equation}
This momentum current is absorbed by the walls, because the fields vanish 
in all space outside the inner part of the wave guide.

   Another effect of the magnetic field $(\!\!~\ref{eq:B1})$ on
the wave guide's walls is the electric current induced in the
$\varphi $-direction. Indeed, let us evaluate the line integral
along the infinitesimal rectangular closed path of fig. 1. Using vector
analysis, Maxwell equations and the boundary condition
$(\!\!~\ref{eq:BOUNDARY2})$, one finds
\begin{equation}
\oint {\bf B\cdot }d{\bf l} = \int curl {\bf B\cdot }d{\bf s} =
\int 4\pi {\bf j\cdot }d{\bf s}.
\label{eq:4PIJ}
\end{equation}
Thus, a nonzero current ${\bf j}$ is induced on the walls, because
only $B_z$ at the inner part of the wave guide
makes a nonvanishing contribution to the line integral.
This outcome proves that a time-dependent (and $z$-dependent) 
electric current flows along the $\varphi $-direction of the
wave guide's walls and that fields of this current
are part of the solution $(\!\!~\ref{eq:E1})$ and $(\!\!~\ref{eq:B1})$.
This electric current sustains the $B_z$ related
standing wave in the radial direction. The walls also counteract
against local electromagnetic pressure.

   The dual solution of example 2 of [7] behaves analogously. Using
the same global factor of 
$(\!\!~\ref{eq:E1})$ and $(\!\!~\ref{eq:B1})$, one finds for this case
\begin{equation}
{\bf B} = \omega J_1(ar)e^{i(bz-\omega t)} {\bf u}_\varphi 
\label{eq:B2}
\end{equation}
\begin{equation}
{\bf E} = b J_1(ar)e^{i(bz-\omega t)} {\bf u}_r +
           ia J_0(ar)e^{i(bz-\omega t)} {\bf u}_z .
\label{eq:E2}
\end{equation}
Hence, the boundary conditions $(\!\!~\ref{eq:BOUNDARY})$ yield
\begin{equation}
J_0(aR) = 0.
\label{eq:BOUNDARY3}
\end{equation}
Since $J_0(ar)$ and $J_1(ar)$ have no common root[11],
a nonvanishing radial electric field exists at
the wave guide's walls. It follows from Maxwell equation
$div {\bf E} = 4\pi \rho $ that a time dependent and $z$-dependent
charge density is built on the inner part of the  wave guide's walls.
Thus, we have also in Example 2 a current that flows on the walls
and affects the fields inside the wave guide.

   Let us examine an analogous experimental setup. Here the source
of the radiation at $z=-L$ is the same as that of the first
experiment but the wave guide is removed. This situation is
different from the wave guide case. Indeed, the fields of a
closed electromagnetic system depend on charges and currents
at the retarded space-time points (see [10], pp. 158-160 or 
[8], p. 225).
Therefore, the wave guide's solutions 
clearly do not hold for this case because here the current along 
the wave guide walls is missing. 

Since in the second experiment
the region at $z=0$ satisfies the wave zone requirements
(see [10] p. 170 or [8], p. 392)
\begin{equation}
L\gg \lambda ,\;\; L\gg 2R,
\label{eq:WAVEZONE}
\end{equation}
one can use the wave zone solution. Let $\bf A$ denote
the retarded vector potential at the wave zone. Thus, one finds 
the fields (see [10] p. 171)
\begin{equation}
{\bf B} = {\bf \dot{A}\times n},
\label{eq:BWZ}
\end{equation}
\begin{equation}
{\bf E} = ({\bf \dot{A}\times n}){\bf \times n}
\label{eq:EWZ}
\end{equation}
where ${\bf n}$ is a unit vector in the radial direction.

   It turns out that the solution for the free space experiment is
inherently different from the one which fits the wave guide's
inner space. In particular, in the case of free space, fields
at the wave zone are perpendicular to the radius vector from the
source to the field point. On the other hand, the wave guide solution 
contains a $z$-component ($B_z$ or $E_z$) which is an inherent part 
of the solution. As shown above, the $B_z$ (or $E_z$) field is associated
with the electric current induced on the wave guide's walls.
This conclusion obviously holds for any pattern of source
elements put at the same spatial region as the one used here,
because the analysis does not refer to the source's details. Thus,
the results disagree with the claim of [9].

   One can use general arguments for proving that a diffraction-free
electromagnetic beam that has a nonvanishing $z$-component for at
least one of the fields, contains transverse standing wave. Indeed,
the beam carries energy and therefore ${\bf S}$ of
$(\!\!~\ref{eq:POYNTING})$ does not vanish. Hence, ${\bf E}$ is
not parallel to ${\bf B}$ and, due to the $z$-component of the
fields, ${\bf S}$ has a nonvanishing transverse component. Now,
the diffraction-free property of the beam prevents energy from
flowing transversally. Hence, the transverse component of ${\bf S}$ 
is a standing wave.

   It can also be proved that all solutions of [7] have a
nonvanishing $z$-component of at least one of the fields. Indeed,
the vector potential $\bf A$ takes the form (see p. 1556 therein)
\begin{equation}
{\bf A} = \sum _{n} (\alpha _n {\bf M}_n + \beta _n {\bf N}_n), 
\label{eq:A}
\end{equation}
where $\alpha _n$ and $\beta _n$ are numerical coefficients of the 
expansion. Here
\begin{equation}
{\bf M}_n = curl [J_n(ar)e^{i(bz + n\varphi - \omega t)}{\bf u_z}]
\label{eq:MN}
\end{equation}
and 
\begin{equation}
{\bf N}_n = \frac {1}{k}curl {\bf M}_n
\label{eq:NN}
\end{equation}
where $k$ is the wave number. Now ${\bf N}_n$ contains a $z$-component
(see p. 1557 therein). Hence, if $\beta _n \neq 0$ then
${\bf E} = -\partial {\bf A}/\partial t = i\omega {\bf A}$ 
has a $z$-component too. In other cases all
$\beta _n = 0$, which mean that for at least one $n$, $\alpha _n \neq 0$.
Here the magnetic field 
${\bf B} = curl {\bf A} = \alpha _n curl{\bf M} = k\alpha _n {\bf N}$, 
which means that $B_z \neq 0$ and the proof is completed. 

   It follows that the family of solutions of [7] involves standing
waves associated with the $z$-components of the solutions. This
diffraction-free family of solutions may fit cylindrical wave guides
but are unsuitable for the case of a free space.

   Example 4 of [7] (see p. 1558) is the last one which is 
analyzed here in detail. This example contains one component which is
proportional to $J_0(ar)$ and is $\varphi $-invariant. Although it
has a $\varphi $-dependent $z$-component term which is
associated with a standing wave, it looks simpler to show another
problem of this solution. The vector potential of this example is given
in Cartesian coordinates 
\begin{equation}
{\bf A} = -i\alpha [aJ_0(ar)\,{\bf u}_x - 
i\frac {a^2}{b}J_1(ar)cos \varphi \,{\bf u}_z]e^{i(bz - \omega t)}.
\label{eq:EX4A}
\end{equation}
Using ${\bf E} = -\partial {\bf A}/\partial t$, one finds
\begin{equation}
{\bf E} = \alpha \omega [aJ_0(ar)\,{\bf u}_x - 
i\frac {a^2}{b}J_1(ar)cos \varphi \,{\bf u}_z]e^{i(bz - \omega t)}.
\label{eq:EX4E}
\end{equation}
Let us examine the $z$-component of the Poynting vector which
represents energy current flowing along the beam's direction, 
namely, the quantity which is analogous
to $(\!\!~\ref{eq:PZ})$ of Example 1.
Examining $(\!\!~\ref{eq:EX4E})$,
one finds that only $B_y$ is needed for this purpose. Thus,
$(curl\,{\bf A})_y$ of $(\!\!~\ref{eq:EX4A})$ is
\begin{equation}
B_y = \alpha [(ab - \frac {a^3}{2b})J_0(ar) +
\frac {a^3}{2b}cos 2\varphi \,J_2(ar)]
e^{i(bz - \omega t)}.
\label{eq:EX4BY}
\end{equation}
Hence, the required $z$-component of the Poynting vector is obtained
as the product of the real parts of $E_x$ of 
$(\!\!~\ref{eq:EX4E})$ and $B_y$ of $(\!\!~\ref{eq:EX4BY})$
\begin{equation}
S_z = \alpha ^2 \omega [(a^2 b - \frac {a^4}{2b})J_0^2(ar) +
\frac {a^4}{2b}cos 2\varphi \,J_0(ar)J_2(ar)]
cos^2(bz - \omega t).
\label{eq:EX4SZ}
\end{equation}

   Let us examine the $z$-component of the energy current near a
point whose radial coordinate is $\bar {R}$ and $J_0(a\bar{R})=0$. In
this neighbourhood $J_2$ is dominant[12] and the contribution of the
$J_0^2(ar)$ term of $(\!\!~\ref{eq:EX4SZ})$ can be ignored. The 
rest of $(\!\!~\ref{eq:EX4SZ})$ is proportional to 
$J_0(ar)J_2(ar)cos\,2\varphi $. 
Now, let us examine the value of $S_z$ on a circle whose radius is
$\bar {R} + \varepsilon $, where $\varepsilon $ is an appropriate
small quantity. Due to the factor $cos\,2\varphi $,
one realizes that $S_z$ takes different signs on this circle.
Hence, in the solution of Example 4 of
[7], energy flows in opposite $z$-directions in certain regions of
space. This property of Example 4 is inconsistent with the notion of 
a beam, where electromagnetic energy flows {\em away} from a
localized source.
 
   It is clear from the analysis carried out above that, 
in free space, one cannot
build a diffraction free {\em beam} from the family of Bessel
functions of [7], because these functions are unsuitable at the
wave zone. 

   Some conclusions can be drawn for the intermediate zone too.
The diffraction free $\varphi $-invariant $J_0(ar)$ function proposed
in [1] does {\em not} belong to the solutions of [7]. Indeed, in
[7], there are only two truly $\varphi $-invariant solutions. They
are the dual solutions of Examples 1 and 2 which are discussed above.
As proved in this work, 
the $z$-component of the energy current is proportional
to $J_1^2(ar)$. Hence, the flow of energy {\em vanishes along the
$z$-axis}. It is also proved above that Example 4 of [7],
where there is one $J_0$ term which is $\varphi $-invariant,
does not describe a beam of electromagnetic radiation
and its $z$-component is not $\varphi $-invariant.
It follows that experiments using a $\varphi$-invariant
setup and showing a strong peak at the center (like [1,3,4]) should
not be interpreted by means of diffraction free solutions.


\newpage
References:
\begin{itemize}
\item[{*}] Email: eli@tauphy.tau.ac.il
\item[{[1]}] J. Durnin, J. J. Miceli, Jr. and J. H. Eberly,
Phys. Rev. Lett. {\bf 58}, 1499 (1987).
\item[{[2]}] Due to this large number of papers, the rather short
reference list presented here cannot be regarded as an adequate
representation of the relevant literature. For having further reference,
readers may use articles mentioned here or a citation database.
\item[{[3]}] C. A. McQueen, J. Arit and K. Dholakia, Am. J. Phys.
{\bf 67}, 912 (1999).
\item[{[4]}] T. Wulle and S. Herminghaus, Phys. Rev. Lett.
{\bf 70}, 1401 (1993).
\item[{[5]}] D. Mugnai, A. Ranfagni and R. Ruggeri, Phys. Rev.
Lett. {\bf 84}, 4830 (2000).
\item[{[6]}] N. P. Bigelow and C. R. Hagen, Phys. Rev.
Lett. {\bf 87}, 059401 (2001);
 H. Ringermacher and L. R. Mead, Phys. Rev.
Lett. {\bf 87}, 059402 (2001);
Thilo Sauter and Fritz Paschke, Phys. Lett. {\bf 285}, 1
(2001).
\item[{[7]}] Z. Bouchal and M. Olivik, J. Mod. Opt. {\bf 42},
1555 (1995).
\item[{[8]}] J. D. Jackson, {\em Classical Electrodynamics} (John Wiley, 
New York,1975). p. 220. 
\item[{[9]}] S. V. Kukhlevsky, G. Nyitray and V. L. Kantsyrev,
Phys. Rev. {\bf E64}, 026603 (2001).
\item[{[10]}] L. D. Landau and E. M. Lifshitz, {\em The Classical
Theory of Fields} (Pergamon, Oxford, 1975). P. 81.
\item[{[11]}] M. Abramowitz and I. Stegun, {\em Handbook of
Mathematical Functions}, (U.S. Government Printing Office, 
Washington, 1972). p. 370.
\item[{[12]}] Due to [11],
the roots $(r > 0)$ of $J_n(r)$ and $J_{n + 1}(r)$ are
simple, do not coincide and interlace. 
Hence, the recurrence formula $2J_1(r)/r=J_0(r)+J_2(r)$ proves
that positive roots of $J_0(r)$ and $J_2(r)$ do not coincide.
\end{itemize}

\newpage
Figure captions: \\

\noindent
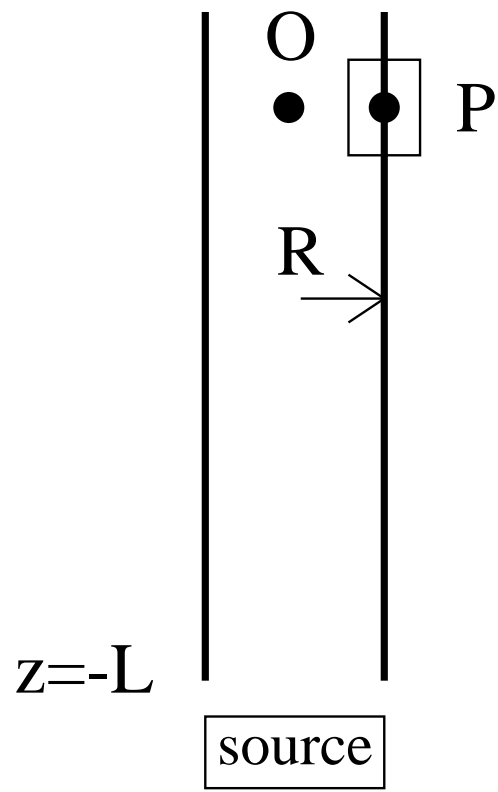
Fig. 1: \\

   Electromagnetic radiation is emitted from a source into a 
cylindrical wave guide whose radius is $R$. The source is at $z=-L$
and $L\gg 2R$. $O$ denotes the origin of coordinates and the 
rectangle at point $P$ denotes a closed integration path (see text).

\end{document}